\documentclass[11pt,a4paper]{article}

\usepackage{tikz}
\usepackage{tikz-feynman}
\tikzfeynmanset{compat=1.1.0}

\usepackage{float} %%% to plave figures here

\usepackage{cite}
\usepackage{amsmath, amsthm, amssymb,slashed}
\usepackage{yfonts}

\usepackage{subcaption}
\usepackage{textcomp}
\usepackage{xfrac}  
\usepackage{fancyhdr}
\usepackage{mathbbol}
\usepackage{url}
\usepackage{color}
\usepackage{bbm}
\usepackage{rsfso}
\usepackage{dsfont}
\usepackage{bm}
\usepackage{amsfonts}
\usepackage{mathrsfs}
\usepackage{slashed}
\usepackage{wasysym}
\usepackage{ytableau}
\usepackage{cancel}
\usepackage[normalem]{ulem}%provides strikeout through \sout{ }

\usepackage{graphicx}

\usepackage{jheppub}

\def\bl{\begin{equation}\begin{aligned}}
\def\el{\end{aligned}\end{equation}}
\def\beal{\begin{align}}
\def\eal{\end{align}}
\def\be{\begin{equation}}
\def\ee{\end{equation}}
\def\bpm{\begin{pmatrix}}
\def\epm{\end{pmatrix}}
\def\bsm{\begin{bmatrix}}
\def\esm{\end{bmatrix}}
\def\bvm{\begin{vmatrix}}
\def\evm{\end{vmatrix}}
\def\bVM{\begin{Vmatrix}}
\def\eVM{\end{Vmatrix}}
\def\bea{\begin{eqnarray}}
\def\eea{\end{eqnarray}}

\def\1{{\bf 1}}
\def\2{{\bf 2}}
\def\3{{\bf 3}}
\def\4{{\bf 4}}

\usepackage{indentfirst}

\newcommand{\OMIT}[1]{}

\usepackage{mathtools}
\usepackage{fontawesome}
\usepackage{microtype,shapepar,xcolor}
\title{The Compact $X$ and $Z$ and their Invisible Molecular Partners}
\date{\today}
\author[a]{A. Carducci}\emailAdd{andrea.carducci@uniroma1.it}
\author[b]{B. Grinstein,}
\author[a]{D. Germani,}\emailAdd{davide.germani@uniroma1.it}
\author[a]{A.D. Polosa}
\affiliation[a]{
Universit\`a degli Studi di Roma La Sapienza and INFN Section of Roma 1, Piazzale Aldo Moro 5, 00185 Roma, Italy
}
\affiliation[b]{
University of California, San Diego, 9500 Gilman Drive, La Jolla, CA 92093, USA
}
\abstract{
We propose a model explaining the strong isospin violations observed in $X(3872)$ decays. The  $X$ is assumed to be a compact isosinglet tetraquark  and its charged partners to  form an isotriplet, together with an additional  neutral state mixing with the isosinglet. The isotriplet results from the loose binding of open charm mesons and,  in isolation, has  strongly suppressed production rates.  However, its neutral component can still mix with the isosinglet    and induce the large isospin violations observed.  The compact-isosinglet/molecular-isotriplet  pattern appears reversed when the  $Z(3900)$ resonances are considered. The $Z$ particles  observed correspond to a compact isotriplet and there is no evidence of a neutral  isosinglet $Z$. We show that no  isospin violation in the $Z$ decays is expected. 
}

\begin{document}
\maketitle
\newpage
\section{Introduction}
The $X(3872)$ exhibits special features that pose challenges to its interpretation. One such peculiarities is its proximity to the $D\bar D^*$ open charm meson threshold. The other is isospin violation observed in its strong decays. Various collaborations have measured the ${\mathcal{I}}$ ratio \cite{Belle:2005lfc,BESIII:2019esk,BaBar:2010wfc}
\begin{equation}
   {\cal I}= \frac{{\text{Br}}(X\to J/\psi\,\pi^+\pi^-\pi^0)}{{\text{Br}}(X\to J/\psi\,\pi^+\pi^-)}=\begin{cases}
        1.0\pm0.4\,\text{(stat)}\pm0.3\,\text{(syst)} & \text{[Belle]}\\
        0.8\pm0.3 & \text{[BABAR]}\\
         1.43^{+0.28}_{-0.23} & \text{[BESIII]}
    \end{cases}\,.
    \label{eq:br_ratioI}
\end{equation}
Assuming that the three-pion final state predominantly arises from the decay $X\to J/\psi\,\omega$, while the two-pion final state stems primarily from $X\to J/\psi\,\rho^0$, this result indicates a significant isospin violation in the strong decay channels of the $X(3872)$.

The LHCb collaboration \cite{LHCb:2022jez} has also measured the ratio of the isospin-violating to isospin-conserving $X(3872)$ couplings
\begin{equation}
    \mathcal{G}=\frac{g_{X(3872)\to J/\psi\rho^0}}{g_{X(3872)\to J/\psi \omega}} = 0.29 \pm 0.04
    \label{eq:grho_gomega}
\end{equation}
and compared it with the case of the $\psi(2S)$, for which
\begin{equation}
    \frac{g_{\psi(2S)\to J/\psi\pi^0}}{g_{\psi(2S)\to J/\psi \eta}} = 0.045 \pm 0.001,
    \label{eq:psi-I-conserve}
\end{equation}
highlighting that a pure charmonium model is disfavored. To explain \eqref{eq:br_ratioI},  Braaten and M. Kusunoki suggested that the isospin violation is actually not as large as it might appear due to the differences in the phase space of the two decays and in the respective couplings for $\rho\to\pi^+\pi^-$  and $\omega\to\pi^+\pi^-\pi^0$ \cite{Braaten:2005ai}. The more recent study result in \eqref{eq:grho_gomega} runs counter to this suggestion. Even earlier, Tornquist proposed  the isospin-breaking mass difference of the neutral and charged $D$ mesons as the origin of isospin breaking in $X$ decays \cite{Tornqvist:2004qy}. To this end, he assumes that the $X(3872)$ is a $D\bar{D}^*$ molecular state,\footnote{Drawing an analogy from the deuteron, Ref.~\cite{Tornqvist:2004qy} assumes that the molecule is bound by a Yukawa potential mediated by pion exchange. However, it was demonstrated in Ref.~\cite{Esposito:2023mxw} that pion exchange does not produce a binding potential because the $D^{*0}$ mass is slightly larger than the sum of the $D^0$ and $\pi^0$ masses. As a result, the potential is not of the Yukawa type.} and further assumes that the molecule is  predominantly formed by the $D^0\bar{D}^{*0}$ component due to the higher mass of the $D^+D^{*-}$, resulting in a superposition of $I=0$ and  $I=1$ states. However, pure molecular models run into difficulties accounting for the rate of prompt $X$ production at colliders \cite{Bignamini:2009sk, Artoisenet:2009wk,
Esposito:2013ada,Esposito:2015fsa,Esposito:2020ywk}.

We propose that the uncommonly large isospin breaking in $X(3872)$ decays arises from near-resonant mixing of two states, an isosinglet, $X_S$, and the neutral component of an isotriplet, $X_T$. More specifically, we propose that $X_S$ is a compact tetraquark state of the diquark-antidiquark type~\cite{Maiani:2004vq},\footnote{See~\cite{Germani:2025mos,Brambilla:2024thx,Brambilla:2026ujo} for a more recent perspective that uses the Born-Oppenheimer approximation to describe this state.} and that $X_T$ is the neutral component of  an isotriplet of loosely bound $D\bar D^*$ molecules. In keeping with Eq.~\eqref{eq:psi-I-conserve} we assume that the couplings of $X_S$ to isotriplet states and of $X_T$ to isosinglet states are negligible. Furthermore, a priori there is no reason to  expect a hierarchy between the isospin preserving couplings of $X_S$ to isosinglet states  and $X_T$ to isotriplet states.

A significant $X_S$ component of the physical $X(3872)$ state justifies its large yield in high transverse momentum collisions at hadron colliders~\cite{Bignamini:2009sk, Artoisenet:2009wk, Esposito:2013ada, Esposito:2015fsa, Esposito:2020ywk}. By contrast, we assume that $D\bar D^*$ loosely bound molecules have negligible production rates in prompt production and $B$ decays\footnote{In our view, the molecular picture is also severely challenged by the observed $J/\psi\, \gamma$ radiative decays~\cite{Grinstein:2024rcu}}, which also justifies the lack of observation of charged partners for the $X(3872)$.

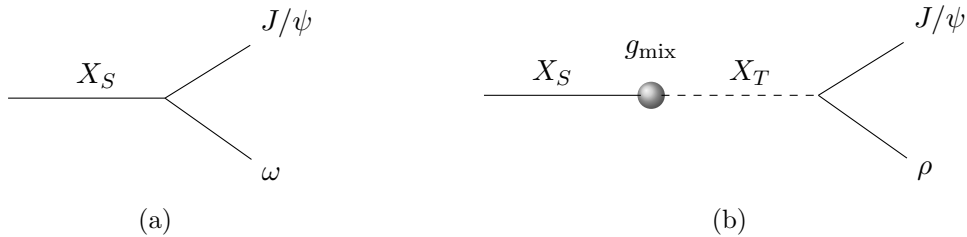
\begin{figure}[H]
\centering

% --- Diagramma (a) ---
\begin{subfigure}[b]{0.45\textwidth}
\centering
\begin{tikzpicture}
  \begin{feynman}
    \vertex (blob) at (2.2,0) {} ;
    \vertex (v1) at (4.4,0) ;
    \vertex at (3.5,0.3) {\(X_S\)};
    \vertex (f1) at (6,1) {$J/\psi$};
    \vertex (f2) at (5.8,-1) {\(\omega\)};

    \diagram*{
      (blob) --  (v1),
      (v1) -- [solid] (f1),
      (v1) -- (f2),
    };
  \end{feynman}
\end{tikzpicture}
\caption*{(a)}
\end{subfigure}
%\hfill
% --- Diagramma (b) ---
\begin{subfigure}[b]{0.45\textwidth}
\centering
\begin{tikzpicture}
  \begin{feynman}
    \vertex (i1) at (0,0);
    \vertex at (0.9,0.3) {\(X_S\)};
%    \vertex [blob, minimum size=1.2cm, fill=gray!20] (blob) at (2.2,0) {};
  \shade[ball color=gray!50] (2.2,0) circle (0.18);
    \vertex at (2.2,0.6) {\(g_{\text{mix}}\)};
    \vertex (v1) at (4.4,0);
    \vertex at (3.5,0.3) {\(X_T\)};
    \vertex (f1) at (6,1) {$J/\psi$};
    \vertex (f2) at (5.8,-1) {\(\rho\)};    
    \diagram*{
      (i1) -- (blob),
      (blob) -- [dashed] (v1),
      (v1) -- [solid] (f1),
      (v1) -- (f2),
    };
  \end{feynman}
\end{tikzpicture}
\caption*{(b)}
\end{subfigure}

\vspace{0.5em}
\caption{Decay mechanisms for \(X(3872) \rightarrow J/\psi\, V\) with \(V = \rho,\, \omega\). (a) Isospin-conserving decay. (b) Isospin-violating decay induced by mixing with the isotriplet.}
\label{fig:X_S_decay}
\end{figure}

Figure \ref{fig:X_S_decay} illustrates our proposal. Only the  $X_S$ is produced (in collisions or $B$-decays). Figure~\ref{fig:X_S_decay}$(a)$ shows its subsequent isospin allowed decay to the isospin singlet state $J/\psi\;\omega$, while Fig.~\ref{fig:X_S_decay}$(b)$ shows it mixing with $X_T$ which then exhibits isospin allowed decay to $J/\psi\;\rho$. In our proposal  the dominant source of isospin violation is in the mixing amplitude, denoted by $g_{\rm mix}$ in Fig.~\ref{fig:X_S_decay}$(b)$. This of course presupposes an isospin violating  mixing amplitude; in Sec.~\ref{sec:mixing} we present a simple model for the origin of mixing and show that the values of the effective mixing parameter $g_{\rm mix}$ necessary to explain the observed isospin violation are well within the reach of that model.

We begin our study, in Sec.~\ref{sec:mixing_XS-decay},  by describing the mixing process and estimating the amplitude for the decay $X_S\to J/\psi\;\omega$, and continue in Sec.~\ref{sec:Gammarho} to propose a theory of $X_T\to J/\psi\;\rho $ decay and estimating its amplitude. An estimate of the mixing amplitude $g_\text{mix}$ can be obtained from the decay $X(3872) \rightarrow D\bar{D}\pi$, as discussed in Sec.~\ref{gmixextr}, and these results are combined in Sec.~\ref{Sec:I-ratio} where we finally give our model's result for the ratio  ${\mathcal I}$ and in Sec.~\ref{Sec:G-ratio} where we use the same ingredients to estimate the ratio $\mathcal{G}$ in \eqref{eq:grho_gomega}. In Sec.~\ref{sec:mixing} we explain how the isospin symmetry-breaking mass splitting in neutral vs. charged $D^{(*)}$ mesons can, though virtual exchange of $D\overline D^*$ pairs, account for the mixing parameter, $g_{\rm mix}$, which, as mentioned above, is a crucial aspect of our proposal. Before summarizing our results and giving an outlook for future progress in Sec.~\ref{conclusions}, we devote Sec.~\ref{sec:z-res} to a discusion of loose ends, such as the role of a triplet of compact tetraquark states and the fate of the charged partners of the $X_T$.

\section{Isospin violation via mixing}
\label{sec:mixing_XS-decay}
In  computing  the effect of mixing as indicated in Fig.~\ref{fig:X_S_decay}(b), we will work in
  the non-relativistic limit, so that the propagators will be taken to be of the form
\begin{equation}
i\Delta(p)=\frac{i}{p^0-m+\frac{{\bm p}^2}{2 m}+i\frac{\Gamma}{2}}= \frac{1}{p^0-m+i\frac{\Gamma}{2}}\,.
\end{equation}
We have written this in terms of the full off-shell energy $p^0$ rather than the customary kinetic
energy $K=p^0-m$, because it is $p^0$ that is conserved at vertices, and in particular in the
$X_S-X_T$ mixing insertion of Fig.~\ref{fig:X_S_decay}(b). On the right hand side we have
specialized to the center of mass (CM) frame. The insertion of $g_{\rm mix}$ in the Feynman graph
corresponds to a contribution to the off diagonal self-energy
\begin{equation}
  i\Pi(p)=i\begin{pmatrix}
    0&g_{\rm mix}\\g_{\rm mix}&0
  \end{pmatrix}
\end{equation}
As usual, the re-summation of self-energy insertions yields,
\[  i\Delta(p)+ i\Delta(p)\, i\Pi(p)\, i\Delta(p) +\cdots=i(\Delta^{-1}(p)+\Pi(p))^{-1} \]
  where
  $$
\Delta^{-1}(p)=\text{diag}(p^0-m_{S}+\tfrac{i}2\Gamma_{S}, p^0-m_{T}+\tfrac{i}2\Gamma_{T})
  $$
After diagonalizing
the $2\times2$ re-summed propagator, the on-resonance ratio of the ST to SS components is
\begin{equation}
  \label{eq:pre-ratio}
\frac{g_{\rm mix}}{\tfrac12\Delta M_X+\tfrac{i}2\Gamma_{T}\pm\sqrt{g_{\rm mix}^2+\tfrac14(\Delta M_X)^2}}  
\end{equation}
at one resonance or the other, which are centered at
\begin{equation}
E_{\rm res,\pm}=\tfrac12(m_{S}+m_{T})\pm\sqrt{g_{\rm mix}^2+\tfrac14(\Delta M_X)^2}\,. 
\end{equation}
We will return below to discussing which of these best models the resonance we associate with the
$X(3872)$. The denominator in Eq.~\eqref{eq:pre-ratio} is conveniently expressed in terms of the
molecular binding energy $B$, so that $m_{T}=m_D+m_{D^*}-B$, and of the distance $E$ between the
$X(3872)$ and threshold,  $E_{\rm  res}=E+m_D+m_{D^*}$, so that
\begin{equation}
\frac{g_{\rm mix}}{\tfrac12\Delta M_X+\tfrac{i}2\Gamma_{T}\pm\sqrt{g_{\rm mix}^2+\tfrac14(\Delta
    M_X)^2}}
=\frac{g_{\rm mix}}{E+B+\tfrac{i}2\Gamma_T}
\end{equation}

The branching ratio in Eq.~\eqref{eq:br_ratioI} is therefore
\begin{equation}
    {\mathcal{I}}=\frac{\Gamma(X\to J/\psi\, \pi^+\pi^-\pi^0)}{\Gamma(X\to J/\psi\, \pi^+\pi^-)}= \bigg|
    \dfrac{g_\text{mix}}{E + B+i\Gamma_T/2}\bigg|^{-2}\,\frac{\alpha\,|G_\omega|^2}{\Gamma(X_T\to J/\psi\pi^+\pi^-)}\,,
    \label{eq:ratio1}
\end{equation}
where the numerical factor $\alpha = 0.645\,\text{MeV}$ arises from
the three-body phase space integral for the decay $\omega \to \pi^+
\pi^- \pi^0$ (see \cite{Braaten:2005ai} (Eq. (33)).\footnote{In Ref.~\cite{Braaten:2005ai}, the corresponding value is $\alpha = 19.4\,\text{keV}$, due to a different normalization convention for the couplings $G_\omega$. The conversion factor between the two conventions is $(m_X / m_\omega)^2$. Furthermore, the neutral isotriplet partner of the $X(3872)$ is not included in their analysis and, as a result, the central factor in Eq.~\eqref{eq:ratio1} (which contains the $X_T$ propagator) is absent in their expression.}
While we will use the right had side of Eq.~\eqref{eq:pre-ratio} extensively, it is worth pointing out that for $\Gamma_T\ll g_{\rm mix}$ and $\Delta M_X\ll g_{\rm mix}$, the ratio is $\approx 1$, indicating resonant amplification of isospin breaking.

Taking into account the experimental uncertainties, the value of $E$ is
\begin{equation}
 E= -0.05 \pm 0.09 \,\text{MeV}
 \label{sette}
\end{equation}
meaning that the resonance can also occur slightly above threshold. We will work with  the central value of $E$ given above.   

As for the value of $\Gamma_T$, we note that  the $D\bar D^*$ pair has $C=+1$ so it cannot annihilate into a virtual photon and the decay with two photon is slower than the $D^*$ decay itself. Also, the $D\bar D^*$ pair has positive parity and therefore  no annihilation at the origin into a virtual pion is possible -- the vertex has to be parity conserving even if the virtual pion itself does not have definite parity.
Therefore we will assume that  $\Gamma_T\simeq \Gamma_{D^{*0}}$. The width of the neutral $D^{*0}$, $\Gamma_{D^{*0}}$, is taken from \cite{Gamma2021}, so that
\begin{equation}
\Gamma_T\simeq \Gamma_{D^{*0}} = 0.0565 \pm 0.0140 \,\text{MeV}\,.
\label{gammaDstar}
\end{equation}

The amplitude $X_S\to J/\psi\,\omega$ in terms of the (dimensionless) coupling $G_{\omega}$ and  $\bm e$ polarizations is \cite{Braaten:2005ai, Maiani:2004vq}
\begin{equation}
    \mathcal{A}_{X_S\, J/\psi\, \omega} = G_\omega\, m_S\, \bm{e}_{X_S}\cdot(\bm{e}_{\omega} \times \bm{e}_{J/\psi})^*\,,
    \label{eq:A_rho}
\end{equation}
giving the partial widths 
\begin{equation}
    \Gamma(X_S \to J/\psi\, \omega) = \frac{p(m_S^2,m_{J/\psi}^2,m_\omega^2)}{8\pi m_X^2} \frac{1}{3}\sum_{\text{pols}}|\mathcal{A}_{X_SJ/\psi\, \omega}|^2
\end{equation}
where $p$ is the decay momentum.

We need now to make an estimate of  $G_\omega$ appearing in Eq.~\eqref{eq:ratio1}.  The  partial width  for the $X_S \rightarrow J/\psi \, \omega$ decay  is 
\begin{equation}
\Gamma(X_S \rightarrow J/\psi\,  \omega) = 2\frac{\langle p \rangle_{\omega}}{8 \pi m_S^2} \cdot (m_S \cdot G_{\omega})^2 
\end{equation}
where
\begin{equation}
\langle p \rangle_{\omega} = \left( \frac{m_{\omega} \Gamma_{\omega}}{\pi} \right)
\int_{(3m_{\pi})^2}^{\infty} ds\,\frac{ p(m_S^2,m_{J/\psi}^2,s)}{(s - m_{\omega}^2)^2 + (m_{\omega} \Gamma_{\omega})^2}\,.
\end{equation}
We use the meson masses \cite{ParticleDataGroup:2024cfk} 
\begin{equation}
    m_{\omega}=782.66 \pm 0.13 \,\text{MeV}, \quad m_{J/\psi}=3096.900 \pm 0.006 \,\text{MeV}\,,
\end{equation}
the width $\Gamma_{\omega}=8.68 \pm 0.13 \,\text{MeV}$ and the branching ratio \cite{ParticleDataGroup:2024cfk} 
\begin{equation}
    \text{Br}(X\rightarrow J/\psi\,\omega) = 0.050 \pm 0.019\,.
\end{equation}
For this estimate we can approximate the $m_S\approx m_X$, since the difference is assumed a very small quantity. The $X$ mass and width\footnote{The width of the $X(3872)$ is taken as twice the imaginary part of the T-matrix pole position~\cite{ParticleDataGroup:2024cfk}.} are
\begin{equation}
m_{X}=3871.64 \pm 0.06 \,\text{MeV}, \quad \Gamma_{X}=0.38 \pm 0.16^{+0.14}_{-0.19}\,\text{MeV}\,.
\end{equation}
The branching ratio $\text{Br}(X\rightarrow J/\psi\,\omega)$ and the  $\Gamma_X$ width are affected by large uncertainties. As we did for the $E$ parameter, we will use their central values.

In this way, we obtain
\begin{equation}
|G_\omega|=0.11 \pm 0.02\,.
\label{eq:G_omega}
\end{equation}

\section{The \texorpdfstring{$X_T$ coupling to the $\rho$}{XT coupling to the rho}}
\label{sec:Gammarho}

\begin{figure}[h]
    \centering
    \includegraphics[width=0.4\linewidth]{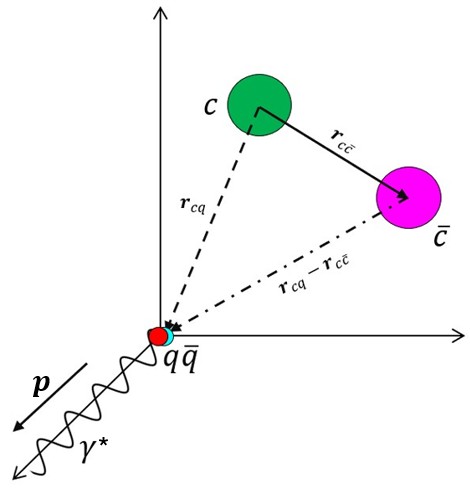}
    \caption{Scheme of the process $X\to J/\psi \gamma^*$. Light quarks annihilate at the origin producing a virtual photon with 4-momentum $p_\mu=\big(\bm{p},i(m_X-E_{J/\psi})\big)$.}
    \label{fig:rad_decay}
\end{figure}

No experimental data are currently available for the process $X_T\to J/\psi\, \pi^+\pi^-$, since the state cannot be produced promptly at high transverse momentum.  
Following Eq.~(30) of Ref.~\cite{Braaten:2005ai}, the differential decay rate into $J/\psi \pi^+\pi^-$ can be expressed as a function of the invariant mass $Q$ of the pion pair as\footnote{The expression below differs from Eq.~(30) of Ref.~\cite{Braaten:2005ai} by a factor $(m_X/m_\rho)^2$, see footnote 3.}
\begin{multline}
\frac{d\Gamma}{dQ}[X \to J/\psi\, \pi^+ \pi^-]
= 
\frac{|G_{\rho}|^2 G_{\rho\pi\pi}^2}{4608 \pi^3 m_X^2 m_\rho^2 m_\psi^2}
\frac{(Q^2 - 4m_\pi^2)^{3/2} \, p(m_X^2,m_{J/\psi}^2,Q^2)}{(Q^2 - m_\rho^2)^2 + m_\rho^2 \Gamma_\rho^2}
\\\times
\left[
(m_X^2 + m_\psi^2)(m_X^2 - m_\psi^2)^2 
- 2(m_X^4 - 4 m_X^2 m_\psi^2 + m_\psi^4) Q^2 
+ (m_X^2 + m_\psi^2) Q^4
\right]
\label{eq:dgammadQ}
\end{multline}
where $G_{\rho\pi\pi}=11.99\pm0.06$~\cite{Braaten:2005ai}.  
We take~\cite{ParticleDataGroup:2024cfk}
\begin{equation}
    m_\rho=775.26\pm0.23\,\text{MeV}, \qquad \Gamma_\rho=147.4\pm0.8\,\text{MeV}\,.
\end{equation}
The dimensionless coupling $G_{\rho}$ is the analogue of $G_{\omega}$ in Eq.~\eqref{eq:A_rho}, and must be estimated in order to compute the total decay width.

We model the decay as dominantly given by the light quark-antiquark, $q\bar{q}$, annihilation into a virtual photon, which subsequently converts into a $\rho^0$ via Vector Meson Dominance (VMD). The effective coupling $G_\rho$ is therefore estimated from the amplitude of the process $X_T\to J/\psi\,\gamma^*\to J/\psi \rho^0$.   
Following Refs.~\cite{Grinstein:2024rcu,Germani:2025mos} the amplitude is determined by the overlap integral between the initial and final states (see Fig.~\ref{fig:rad_decay}),  
\begin{equation}
    I(p)=\int_{\bm{r}_{c\bar{c}},\bm{r}_{cq}} 
    e^{-i\bm p\cdot\left(\frac{\bm{r}_{c\bar{c}}}{2}-\bm{r}_{cq}\right)}\,
    \psi(|\bm{r}_{c\bar{c}}|)\Psi_{\text{mol}}(|\bm{r}_{c\bar{c}}|)
    \psi_q(|\bm{r}_{cq}|)\psi_{\bar{q}}(|\bm{r}_{cq}-\bm{r}_{c\bar{c}}|)
\label{eq:A}
\end{equation}
where the exponential factor accounts for the recoil of the final $J/\psi$, described by the wave function $\psi(r_{c\bar{c}})$ computed as in Refs.~\cite{Grinstein:2024rcu,Germani:2025mos}.  
The molecular $X_T$ is described by the universal shallow bound state wave function~\cite{Braaten:2003he,jackiw},
\begin{equation}
    \Psi_{\text{mol}}(r_{c\bar{c}})=
    \left(\frac{2mB}{4\pi^2}\right)^{1/4}
    \frac{e^{-r_{c\bar{c}}\sqrt{2mB}}}{r_{c\bar{c}}}\,.
\end{equation}
For the light-quark orbitals inside the $D$ and $D^*$ mesons, we employ the Isgur–Scora–Grinstein–Wise (ISGW) wave functions~\cite{Isgur:1988gb},
\begin{equation}
    \psi_{q(\bar{q})}(r)=\frac{b^{3/2}}{\pi^{3/4}}\,e^{-\frac{1}{2}b^2r^2}
\end{equation}
with $b=0.35\,\text{GeV}$~\cite{Isgur:1988gb,Grinstein:2024rcu}.  
The integral depends on the decay momentum
\begin{equation}
  p=|\bm{p}|=p(m_T^2,m_{J/\psi}^2,Q^2)\approx p(m_X^2,m_{J/\psi}^2,Q^2)\,,  
\end{equation}
which is a function of $Q$. We therefore make this dependence explicit from now on, writing $I(Q)$.

This integral must be multiplied by~\cite{OConnell:1995nse}
\begin{equation}
    -ig_{\rho\gamma}=-ie\,\frac{m_\rho^2}{g_\rho}
\end{equation}
where $g_\rho=g_{\rho\pi\pi}=5.96$ according to the VMD model \cite{OConnell:1995nse,PhysRevLett.21.244,Schildknecht:2005xr}. The $q\bar{q}$ annihilation contributes with a factor
\begin{equation}
    -ie\cdot\frac{1}{\sqrt{2}}\cdot 2\left(\frac{2}{3}\right)
    = -i\frac{2\sqrt{2}}{3}e
\end{equation}
where the factors $1/\sqrt{2}$ and $2$ arise from the wavefunction of the $X_T$ state, treated as a bound state of $D^0\bar{D}^{*0}$ and $\bar{D}^0D^{*0}$.  
Finally, the virtual photon propagator reads\footnote{We consider only the scalar part, since all polarizations are already included in Eq.~\eqref{eq:dgammadQ}.}
\begin{equation}
    \frac{1}{-p_0^2+|\bm{p}|^2}
    =\frac{1}{-\left(m_X-\sqrt{m_{J/\psi}^2+|\bm{p}|^2}\right)^2+|\bm{p}|^2}
    =-\frac{1}{m_X^2+m_{J/\psi}^2-2m_X\sqrt{m_{J/\psi}^2+|\bm{p}|^2}}\,.
\end{equation}

\begin{figure}[t]
    \centering
    \includegraphics[width=0.5\linewidth]{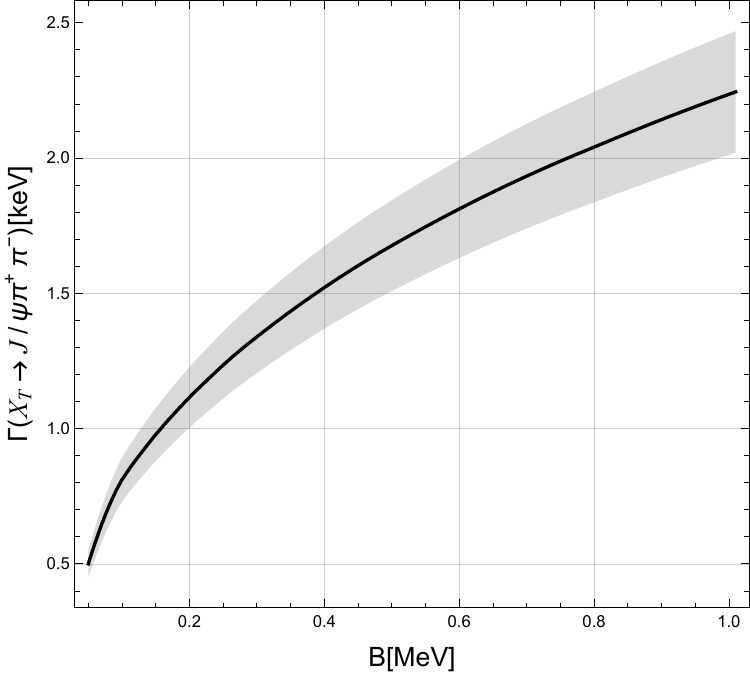}
    \caption{$\Gamma(X_T\to J/\psi \pi^+\pi^-)$ as a function of the binding energy $B$. The gray band represents the $1\sigma$ uncertainty range.}
    \label{fig:Gammarho}
\end{figure}

All contributions together give ($\alpha_{\text{em}}=e^2/4\pi$)
\begin{equation}
    G_\rho(Q)=
    \frac{8\sqrt{2}\pi\alpha_{\text{em}}\,I(Q)}{3g_\rho}
    \frac{m_\rho^2}
    {m_X^2+m_{J/\psi}^2-2m_X\sqrt{m_{J/\psi}^2+p(m_X^2,m_{J/\psi}^2,Q^2)^2}}
    \label{eq:G_rho(Q)}
\end{equation}

Substituting this expression into Eq.~\eqref{eq:dgammadQ} and integrating between 
$Q_{\text{min}}=2m_\pi$ and $Q_{\text{max}}=m_X-m_{J/\psi}$, we obtain the dependence on the binding energy $B$ shown in Fig.~\ref{fig:Gammarho}.

\section{\texorpdfstring{The mixing amplitude from $X(3872)\to D \bar D \pi $ decay}{The mixing amplitude from X(3872) to D bar D pi decay}}
\label{gmixextr}
In order to make a theoretical prediction of the ratio ${\cal I}$ as given in equation~\eqref{eq:ratio1}, we need an empirical determination of $g_{\rm mix}$. 

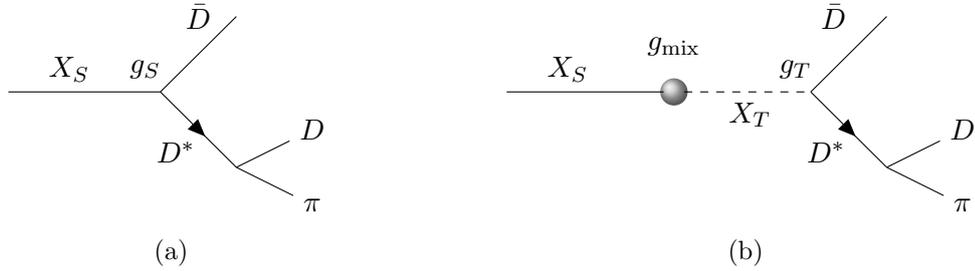
\begin{figure}[h]
\centering
\begin{subfigure}[b]{0.45\textwidth}
\centering
\begin{tikzpicture}
  \begin{feynman}
    \vertex (i1) at (0,0);
    \vertex at (0.8,0.3) {\(X_S\)};
    \vertex (v1) at (2,0);
    \vertex at (1.8,0.3) {\(g_S\)};
    \vertex (d1) at (3,1);
    \vertex at (2.5,1) {\(\bar D\)};
    \vertex (dstar) at (3,-1);
    \vertex at (2.2,-0.8) {\(D^*\)};
    \vertex (pi1) at (4,-1.5) {\(\pi\)};
    \vertex (d2) at (4,-0.5) {\(D\)};
    \diagram*{
      (i1) -- (v1),
      (v1) -- (d1),
      (v1) -- [fermion] (dstar),
      (dstar) -- (d2),
      (dstar) -- (pi1),
    };
  \end{feynman}
\end{tikzpicture}
\caption*{(a)}
\end{subfigure}
%\hfill
% --- Diagramma (b) ---
\begin{subfigure}[b]{0.45\textwidth}
\centering
\begin{tikzpicture}
  \begin{feynman}
    \vertex (i1) at (0,0);
    \vertex at (0.8,0.3) {\(X_S\)};
 %   \vertex [blob, minimum size=1.2cm, fill=gray!20] (blob) at (1.8,0) {};
   \shade[ball color=gray!50] (2.2,0) circle (0.18);
    \vertex at (2.2,0.6) {\(g_{\text{mix}}\)};
    \vertex (v1) at (4,0);
    \vertex at (3.8,0.3) {\(g_T\)};
    \vertex at (3.2,-0.3) {\(X_T\)};
    \vertex (d1) at (5,1);
    \vertex at (4.3,1) {\(\bar D\)};
    \vertex (dstar) at (5,-1);
    \vertex at (4.2,-0.8) {\(D^*\)};
    \vertex (d2) at (6,-0.5) {\(D\)};
    \vertex (pi1) at (6,-1.5) {\(\pi\)};

    \diagram*{
      (i1) -- (blob),
      (blob) -- [dashed] (v1),
      (v1) -- (d1),
      (v1) -- [fermion] (dstar),
      (dstar) -- (d2),
      (dstar) -- (pi1),
    };
  \end{feynman}
\end{tikzpicture}
\caption*{(b)}
\end{subfigure}

\vspace{0.5em}
\caption{Decay diagrams for $X_S\to D\bar{D}\pi$. (a) Direct decay. (b) Decay via isospin mixing through \( X_T \).}
\label{fig:ddstar-decay}
\end{figure}

A calculation of the mixing amplitude  $g_\text{mix}$ can be obtained from the decay $X(3872) \rightarrow D\bar{D}\pi$.  
Again we use the observation that the $X_S$ component is the only one produced at high rate in prompt production or $B$ decays. This proceeds via a $\bar D$ meson and a virtual $D^*$, which subsequently decays into $D\pi$ (see Fig \ref{fig:ddstar-decay}(a)), or via a mixing into the $X_T$ molecule which eventually decays into  $D\bar D\pi$ as in Fig.~\ref{fig:ddstar-decay}(b). The differential decay rate is\footnote{The following masses are used \cite{ParticleDataGroup:2024cfk} 
\begin{equation}
    m_D = 1864.84 \pm 0.05~\text{MeV},\quad m_{D^*} = 2006.85 \pm 0.05~\text{MeV},  \quad m_{\pi} = 134.9768 \pm 0.0005~\text{MeV}
\end{equation}
}~\cite{Esposito:2025hlp}
\begin{equation}
\frac{d\Gamma_{\text{th}}(X \to D\bar{D}\pi)}{ds} = 
2g^2 \frac{p(m_X^2, m_D^2, s)}{8\pi m_X^2} 
\frac{1}{\pi}
\frac{ \frac{s}{m_{D^*}} \text{Br}(D^* \to D\pi) \Gamma_{D^*}}{(s - m_{D^*}^2)^2 + \left(  \frac{s}{m_{D^*}} \Gamma_{D^*} \right)^2}
\frac{\frac{p(s, m_D^2, m_\pi^2)}{2\sqrt{s}}}{\frac{p(m_{D^*}^2, m_D^2, m_\pi^2)}{2m_{D^*}}}
\label{eq:diff_XDDpi}
\end{equation}
where 
\begin{equation}
    g^2 = 8mm_X^2 
\left| 
  g_S -\frac{g_\text{mix} \, g_T }{E + B+i\Gamma_T/2} 
\right|^2\,.
\label{eq:couplingd}
\end{equation}
and $m$ is the reduced mass of the $D^0\bar{D}^{*0}$ pair. 
The coupling $g_T$ of the loosely bound $X_T$ to its components is approximately given by\footnote{The approximation assumes that the binding energy is small, and that the bound state has vanishing orbital angular momentum (s-wave). See also Ref.~\cite{Weinberg:1965zz}.} \cite{landau}  
\begin{equation}
    g_T^2=\frac{2\pi}{m}\sqrt{\frac{2B}{m}}\,.
    \label{eq:landau}
\end{equation}
$g_S$ is the coupling of the compact tetraquark to the $DD^*$ pair, defined in the non-relativistic interaction Lagrangian\footnote{The fields $\Phi_i$ that appear in the non-relativistic Lagrangian have energy dimension $[\Phi_i]=[E]^{3/2}$, so the coupling $g_S$ has dimension $[g_S]=[E]^{-1/2}$.} (see Eq.~(192) of Ref.~\cite{Esposito:2025hlp}),
\begin{equation}
    \mathcal{L}_{\text{int}}
    = -\frac{g_S}{\sqrt{2}}\,X_S^\dagger (D^0\bar{D}^{*0})
      + \frac{g_S}{\sqrt{2}}\,X_S^\dagger(D^{+}D^{*-})
      + \text{c.c.} + \text{h.c.}\,.
\end{equation}
Because there is at present no \textit{ab inito} calculation of $g_S$, its numerical value must be extracted from the scattering data of the $X(3872)$. Unfortunately, the LHCb analysis currently available does not include the isotriplet component~\cite{LHCb:2020xds}, since the Flatté model used to fit the data accounts for scattering through a single isosinglet resonance, and the subsequent decay into $J/\psi\pi^+\pi^-$ is encoded in the factor $\Gamma_\rho(E)$ (see Eq.~(4) of Ref.~\cite{LHCb:2020xds}), where $E$ denotes the energy measured from the $D^0D^{*0}$ threshold. 
Hence, the fitted coupling
\begin{equation}
    g_{\text{LHCb}}=0.108\pm0.003,
\end{equation} 
cannot be directly trace back to $g_S$, since the states $X_S$ and $X_T$ mix, so the data should be fitted accordingly in order to properly disentangle the different components.\footnote{Regarding the fitted width $\Gamma_\rho(E_X)$, the analysis reports a value of order $\mathcal{O}(100\,\text{keV})$.  
This result is not consistent with the value reported by the PDG~\cite{ParticleDataGroup:2024cfk}, which quotes a branching ratio $\text{Br}(X\to J/\psi \pi^+\pi^-)=(4.3\pm1.4)\%$, even assuming a total width $\Gamma_X\sim\mathcal{O}(1\,\text{MeV})$ as extracted from the fit.  
It should nevertheless be noted that $\Gamma_\rho(E_X)$ depends on the coupling $f_\rho$, which is affected by a $50\%$ uncertainty.
}
In the absence of such an analysis, for now we will use the LHCb value as a crude estimate, which is nevertheless sufficient for the purposes of the present study. We will return to this point once a complete fit becomes available. The coupling $g_{\text{LHCb}}$ is related to our definition of $g_S$ through the relation (see Eq.~(226) in Ref.~\cite{Esposito:2025hlp})
\begin{equation}
    g_S=\sqrt{\frac{\pi}{m}\,g_{\text{LHCb}}}
    =(1.87 \pm 0.03)\times 10^{-2}\,\text{MeV}^{-1/2}.
 \label{eq:gc}
\end{equation}
In Sec.~\ref{sec:mixing} we will show that this result is nevertheless in the same ballpark as the value predicted by our mixing model.

\begin{figure}[H]
    \centering
    \begin{subfigure}{0.48\linewidth}
        \centering
        \includegraphics[width=\linewidth]{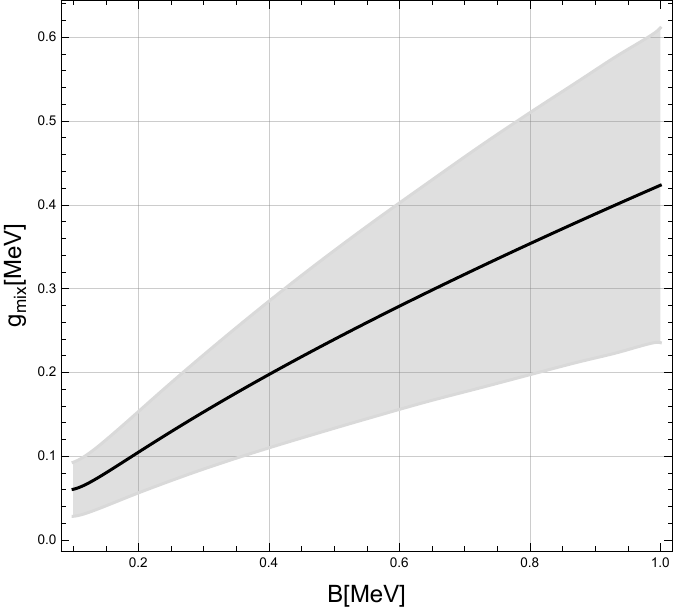}
    \caption{}    
    \label{fig:gmix1}
    \end{subfigure}
    \hfill
    \begin{subfigure}{0.48\linewidth}
        \centering
        \includegraphics[width=\linewidth]{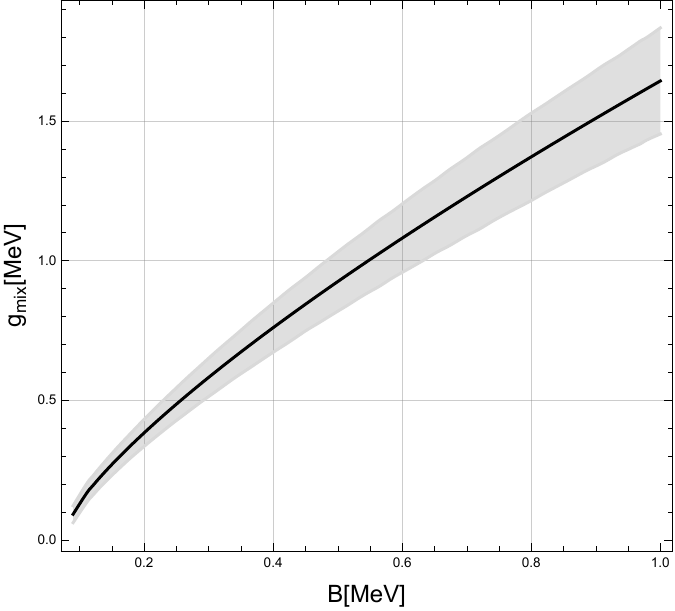}
    \caption{}    
    \label{fig:gmix2}
    \end{subfigure}
    \caption{$g_{\text{mix}}$ as a function of $B$ for the two solutions found. The black line represents the central value of $g_{\text{mix}}$ and the gray region contains the values within one sigma.}
    \label{fig:gmix12}
\end{figure}

Integrating \eqref{eq:diff_XDDpi} from $s_{\text{min}} = (m_D + m_\pi)^2$ to $s_{\text{max}} = (m_X - m_D)^2$, we get the partial decay rate as a function of the phenomenological coupling $g_\text{mix}$. This has to be compared with the experimental partial decay rate
\begin{equation}
    \Gamma_{\text{exp}}(X \to D\bar{D}\pi) = \Gamma_X\,\text{Br}(X\rightarrow D\bar{D}\pi)\,.
\end{equation}
By solving the equation $\Gamma_{\rm th} = \Gamma_{\rm exp}$ for $g_{\text{mix}}$ as a function of $B$, using the current determinations of the branching fractions \cite{ParticleDataGroup:2024cfk}
\begin{equation}
    \text{Br}(X \to D\bar{D}\pi) = 0.55 \pm 0.28,\quad\text{Br}(D^*\to D\pi) = 0.647 \pm 0.009\,,
\end{equation}
we obtain the two solutions shown in Fig.~\ref{fig:gmix12}, since the coupling enters through a squared modulus. We displays the results in the range of $B$ where real solutions for $g_{\text{mix}}$ exist, namely for $B \geq 0.09\,\text{MeV}$.

%{\bf{\emph {The $\bm {\mathcal I}$ ratio.}}} 
\section{Determination of the \texorpdfstring{${\mathcal I}$}{I} ratio}
\label{Sec:I-ratio}
\begin{figure}[H]
        \centering
        \includegraphics[width=0.6\linewidth]{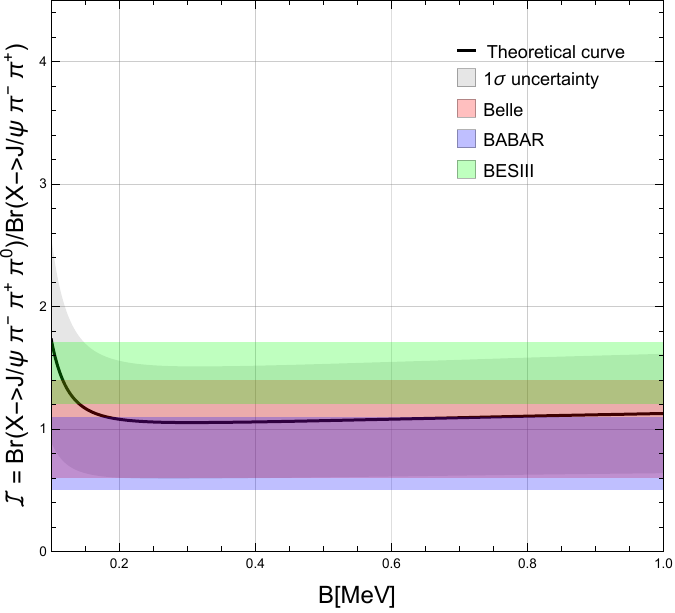}
    \caption{$\mathcal{I}$ as a function of $B$ for the solution $g_\text{mix}$ shown in Fig.~\ref{fig:gmix2}. The black line represents the central value of $\mathcal{I}$ obtained from Eq.~\eqref{eq:ratio_fin}, with the $1\sigma$ uncertainty band shown in gray. The horizontal bands correspond to the experimental values in Eq.~\eqref{eq:br_ratioI}.}
    \label{fig:iso12}
\end{figure}

To test the validity of our  model,  we are now in the position to compute ${\cal I}$ from Eq.~\eqref{eq:ratio1}
\begin{equation}
    {\mathcal{I}}=\frac{\text{Br}(X\to J/\psi\, \pi^+\pi^-\pi^0)}{\text{Br}(X\to J/\psi\, \pi^+\pi^-)}= \bigg|
    \dfrac{g_\text{mix}}{E + B+i\Gamma_T/2}\bigg|^{-2}\,\frac{\alpha\,|G_\omega|^2}{\Gamma(X_T\to J/\psi\pi^+\pi^-)}\,.
\label{eq:ratio_fin}    
\end{equation}
Using the values of $\Gamma_{D^*}$ and $G_\omega$ given in Eqs.~\eqref{gammaDstar} and~\eqref{eq:G_omega}, respectively, together with the width $\Gamma(X_T\to J/\psi \pi^+\pi^-)$ discussed in Sec.~\ref{sec:Gammarho} and the coupling $g_{\rm mix}$ discussed above, we get the results shown in Fig.~\ref{fig:iso12} as a function of the $B$ parameter. We see that the value of $g_{\text{mix}}$ shown in Fig.~\ref{fig:gmix2} yields an excellent agreement with the measured ${\cal I}$, so we show only this solution. In Sec.~\ref{sec:mixing}, we will study a possible theoretical basis for the $g_{\rm mix}$ coupling, providing a motivation for choosing the $g_{\rm mix}$ solution given in Figure \ref{fig:gmix2}  as the physical one.

\section{Determination of \texorpdfstring{$\mathcal{G}$}{G} ratio}
\label{Sec:G-ratio}
In addition to the ratio $\mathcal{I}$, we can also provide an estimate for $\mathcal{G}$ in \eqref{eq:grho_gomega}. 
In the notation adopted throughout this work, we identify 
$g_{X(3872)\to J/\psi \omega}$ with $G_\omega$ in Eq.~\eqref{eq:G_omega}.  
The isospin-violating coupling $g_{X(3872)\to J/\psi\rho^0}$ is instead given by
\begin{equation}
    g_{X(3872)\to J/\psi\rho^0}
    = \left|\frac{g_{\text{mix}}}{E+B+i\Gamma_T/2}\right|\, G_{\rho}\,.
\end{equation}

We note that $G_\rho$ is not a constant coupling, since it is derived from the VMD model 
(see Eq.~\eqref{eq:G_rho(Q)}).   This can be circumvented by defining a constant coupling
\begin{equation}
    \overline{G}_\rho
    = \sqrt{\frac{\Gamma(X_T\to J/\psi \pi^+\pi^-)}{\mathcal{N}}}
\label{eq:Gbar_rho}
\end{equation}
where
\begin{equation}
    \mathcal{N}
    = \int_{4m_\pi^2}^{(m_X - m_{J/\psi})^2}
      \frac{1}{|G_\rho(Q)|^2}\,
      \frac{d\Gamma}{dQ}[X_T\to J/\psi\pi^+\pi^-]\, dQ\,.
\end{equation}

\begin{figure}[h]
    \centering
    \includegraphics[width=0.5\linewidth]{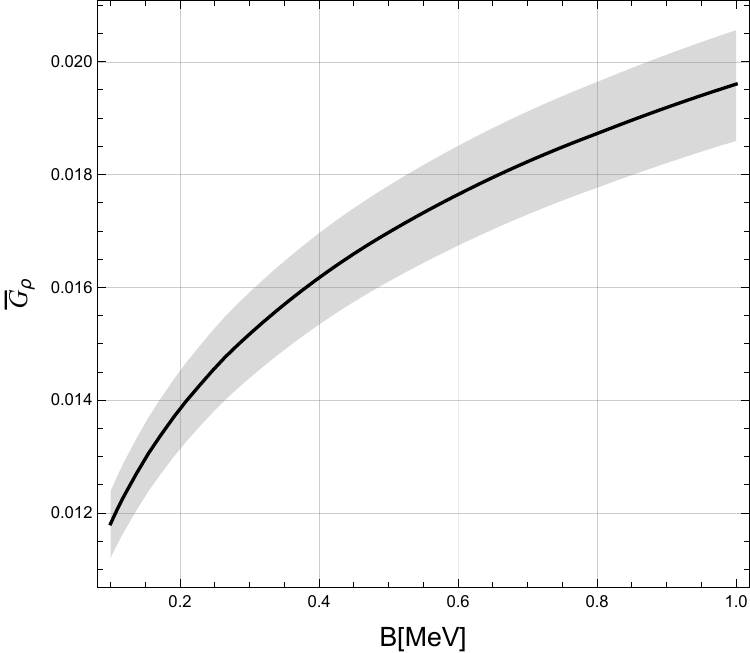}
    \caption{The black line represents the central value of $\bar{G}_\rho$ in Eq.~\eqref{eq:Gbar_rho}, with the $1\sigma$ uncertainty band shown in gray.}
    \label{fig:gXJpsirho}
\end{figure}

Essentially, $\overline{G}_{\rho}$ is the constant coupling that reproduces for $X_T$ the same decay width into $J/\psi \pi^+\pi^-$ as that obtained using 
the full $G_\rho(Q)$. The dependence of $\overline{G}_\rho$ on $B$ is shown in Fig.~\ref{fig:gXJpsirho}.

\begin{figure}[b]
    \centering
    \includegraphics[width=0.55\linewidth]{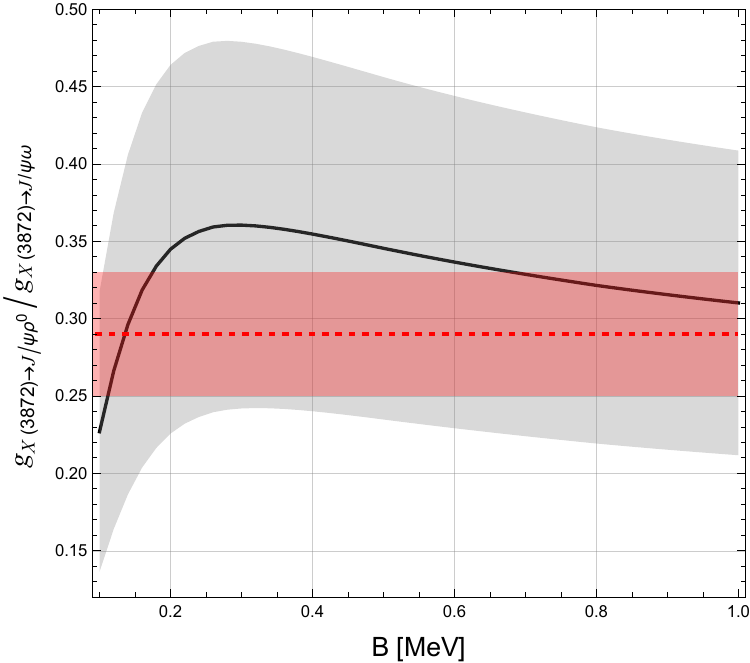}
    \caption{Estimate of the ratio $\mathcal{G}$ defined in Eq.~\eqref{eq:grho_gomega} (the $1\sigma$ uncertainty band is shown in gray). The experimental value from~\cite{LHCb:2022jez} is shown in red.}
    \label{fig:ratiogLHCb}
\end{figure}

The ratio in Eq.~\eqref{eq:grho_gomega} can then be estimated as
\begin{equation}
    \mathcal{G}
    = \frac{g_{X(3872)\to J/\psi\rho^0}}{g_{X(3872)\to J/\psi \omega}}
    = \left|\frac{g_{\text{mix}}}{E+B+i\Gamma_T/2}\right|
      \frac{\overline{G}_\rho}{G_{\omega}}\,,
\end{equation}
and is displayed in Fig.~\ref{fig:ratiogLHCb} together with the experimental
measurement. Also in this case, we find good agreement between the two using the solution for $g_{\text{mix}}$ in Fig.~\ref{fig:gmix2}.

\section{The mixing model}
\label{sec:mixing}
We will assume that the  mixing  amplitude $g_{\rm mix}$ results from the quantum fluctuation of an $X_S$ into a $D\bar D^*$ meson loop which might convert into  $X_T$. The coupling of $X_S$ to $D$ mesons is $g_S$, as given in~Eq.~\eqref{eq:gc}. The coupling of $X_T$ to $D^0\bar D^{*0}$ is the Landau coupling $g_T$ in Eq.~\eqref{eq:landau}, as long as the $D^0$ and $D^{*0}$ do not have a too large relative recoiling energy, which would drive us outside the assumptions which are at the basis for the derivation of the  $g_T$ formula.\footnote{A charged $D^+D^{*-}$ component in the molecular composition  requires a larger binding energy $B$ and the Landau coupling formula in~\eqref{eq:landau} would be approximate up to neglected  $B^2$ terms, which have  a size of $\sim100$~MeV, still much lower than the masses of $D,D^*$ mesons. }
To fulfill those assumptions, we are compelled to introduce a physical cutoff $\Lambda$ in the computation of the loop. 
%: \ac{qui secondo il ragionamento fatto sopra, invece di E deve essere $E_T = E + B$}
\begin{equation}   
\begin{aligned}
\begin{tikzpicture}[baseline=-0.5ex]
 
  \draw[thick] (-1.5, 0) -- (-0.7,0);
  \node[above] at (-1.1, 0) {$E$};  % Testo sopra la linea sinistra
  
  \draw[thick] (0,0) circle (0.7);

  \node[above] at (0.0,0.7) {$E/2+\ell_0$};
  \node[below] at (0,-0.7) {$E/2-\ell_0$};

  \draw[thick] (0.7,0) -- (1.5,0);
  \node[above] at (1.1, 0) {$E$};
\end{tikzpicture}  &= \int \frac{d^3\ell \, d\ell_0}{(2\pi)^4} \frac{i}{\frac{\ell^2}{2m_D} - (E/2 + \ell_0) -i\epsilon} \cdot \frac{i}{\frac{\ell^2}{2m_{D^*}} - (E/2 - \ell_0) -i\epsilon}= \\
&=- 2mi \int \frac{d^3\ell}{(2\pi)^3} \frac{1}{\ell^2 - 2mE - i\epsilon}=\\
&= i\frac{m}{\pi^2} \left(-\Lambda+\sqrt{-2mE} \arctan\left(\frac{\Lambda}{\sqrt{-2mE}}\right)\right)\,.
\end{aligned}
\label{eq:loopEE}
\end{equation}
where $m$ stands for the reduced mass, $1/m=1/m_D+1/m_{D^*}$ and 
\begin{equation}
 E=m_X-m_{D^*}-m_{D}   
\end{equation}
as in Eq. \eqref{sette}. For the charged $D^+D^{*-}$ loop, we should substitute $E\to E-\Delta$ where $\Delta = m_{D^{*+}}+m_{D^+}-m_{D^{*0}}-m_{D^0} \simeq 8\,\text{MeV}$. The cutoff $\Lambda$ represents the maximum energy scale at which the Landau formula for the coupling is still valid and we can use it to evaluate the diagrams given above.

\begin{figure}[t]
    \centering
    \begin{subfigure}{0.48\linewidth}
        \centering
        \includegraphics[width=\linewidth]{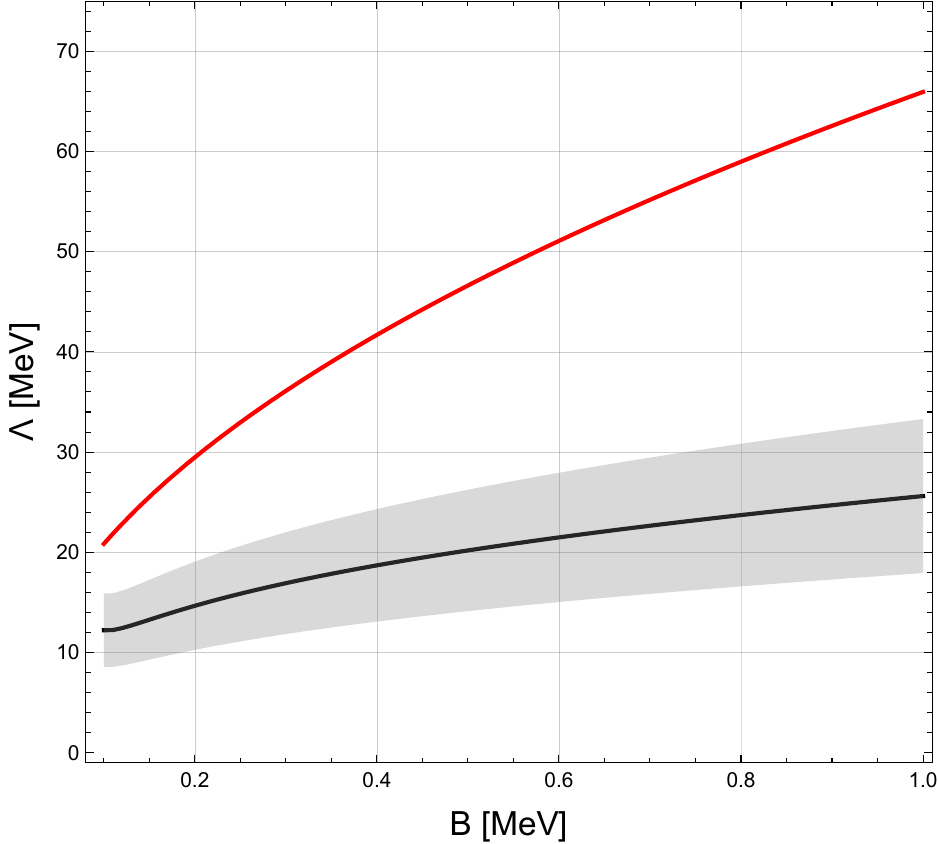}
    \caption{}    
    \label{fig:cutoff1}
    \end{subfigure}
    \hfill
    \begin{subfigure}{0.48\linewidth}
        \centering
        \includegraphics[width=\linewidth]{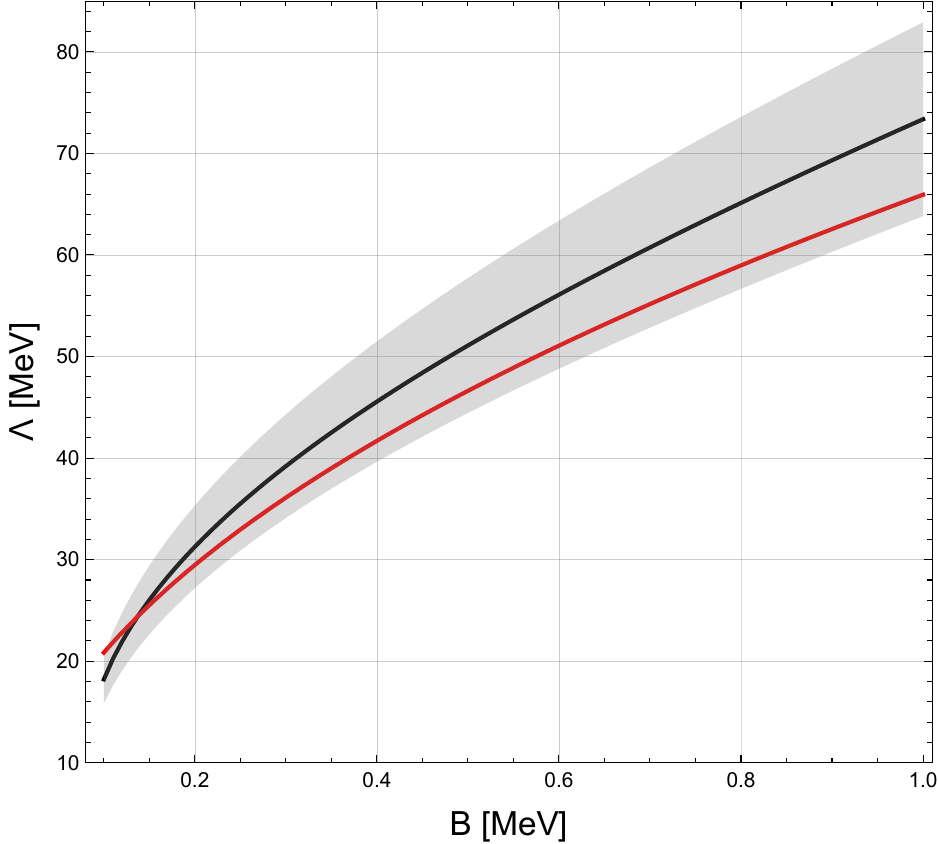}
    \caption{}    
    \label{fig:cutoff2}
    \end{subfigure}
    \caption{$\Lambda$ as a function of  $B$ for the two solutions of $g_\text{mix}$. The black line represents the central value of $\Lambda$ and the gray region contains the values within one sigma. The red line marks the values obtained through the estimate in Eq.~\eqref{eq:cutoff}.}
    \label{fig:cutoff12}
\end{figure}

We are absorbing in the definition of $g_{\rm mix}$  the $1/(2\pi)^4 $ factor coming with the propagator, therefore, using the Feynman rules for the coupling of $X_S$ and $X_T$ as defined in~\cite{Esposito:2025hlp}\footnote{Due to isospin conservation the isosinglet and isotriplet couple to the neutral mesons with a relative minus sign (see \cite{Esposito:2025hlp} Eqs. (188)-(189)).}, we have\footnote{Charged-conjugate diagrams are implicitly included. Since strong interactions preserve $C$ parity, both $X_S$ and $X_T^0$ couple to the $C$-even neutral $D\bar{D}^*$ currents, which introduces a factor $1/\sqrt{2}$ at each vertex, thereby canceling the factor of $2$ arising from the sum of each diagram with its charged-conjugate counterpart.}
\begin{equation} 
\begin{aligned}
(-ig_{\text{mix}}) &= 
\begin{tikzpicture}[baseline=-0.5ex]
 
  \draw[thick] (-2.0, 0) -- (-1,0);
  \node[above] at (-1.5, 0) {$X_S$};  % Testo sopra la linea sinistra
  
  \draw[thick] (0,0) circle (1);

  \node[above] at (0,1.1) {$D^0$};
  \node[below] at (0,-1.1) {$\bar{D}^{0*}$};

  \draw[thick] (1,0) -- (2,0);
  \node[above] at (1.5, 0) {$X_T$};
\end{tikzpicture}
+ 
\begin{tikzpicture}[baseline=-0.5ex]
 
  \draw[thick] (-2.0, 0) -- (-1,0);
  \node[above] at (-1.5, 0) {$X_S$};  % Testo sopra la linea sinistra
  
  \draw[thick] (0,0) circle (1);

  \node[above] at (0,1.1) {$D^+$};
  \node[below] at (0,-1.1) {${D}^{-*}$};

  \draw[thick] (1,0) -- (2,0);
  \node[above] at (1.5, 0) {$X_T$};
\end{tikzpicture} =\\
&= (-ig_S) (i g_T) i\frac{m}{\pi^2} \left(-\Lambda+\sqrt{-2mE} \arctan\left(\frac{\Lambda}{\sqrt{-2mE}}\right)\right)+\\
&  +(ig_S) (i g_T) i\frac{m_+}{\pi^2} \left(-\Lambda+\sqrt{-2m(E-\Delta)} \arctan\left(\frac{\Lambda}{\sqrt{-2m(E-\Delta)}}\right)\right)\,.\\
\end{aligned}
\end{equation}
where $m_+$ is the reduced mass of the $D^+D^{-*}$ mesons.
Since $m_+\simeq m$, we obtain the following expression for the mixing amplitude
\begin{equation}
    g_{\text{mix}} = g_S g_T  \frac{m}{\pi^2} \left[
 \text{arctan} \left( \frac{\Lambda}{\sqrt{-2m(E - \Delta)}} \right)\sqrt{-2m(E-\Delta)} -\text{arctan} \left( \frac{\Lambda}{\sqrt{-2mE}} \right)\sqrt{-2mE} \right]
 \label{eq:gmixcharged}
\end{equation}
which gives $g_{\rm mix}=0$ in the isospin symmetric case $\Delta=0$. Using the dependence of $g_\text{mix}$ on  $B$ (see Figure \ref{fig:gmix12}) in the range of $B$ shown in Figure \ref{fig:iso12}, we estimate $\Lambda$ as a function of $B$. We compare these values with the estimate \cite{Bignamini:2009sk}
\begin{equation}
\Lambda \simeq k_{\rm max}\sim \langle k\rangle+\Delta k=\sqrt{2 m B}+\Delta k = \frac{3}{2}\sqrt{2 m B}
\label{eq:cutoff}
\end{equation}
where $\Delta k$ is found by the minimum uncertainty relation $R_{\rm mol}\cdot \Delta k=1/2$ with $R_{\rm mol}=1/\sqrt{2mB}$. \\
The results are illustrated in Figure \ref{fig:cutoff12}.
We notice that the solution in Figure \ref{fig:cutoff2} is the  one that is compatible with the estimate in Eq.~\eqref{eq:cutoff}. The corresponding solution of $g_\text{mix}$, namely the one shown in Figure \ref{fig:gmix2}, is the one used in Figure \ref{fig:iso12} in which we found a good compatibility with experimental data.  

\begin{figure}[t]
    \centering
    \begin{subfigure}{0.5\linewidth}
        \centering
        \includegraphics[width=\linewidth]{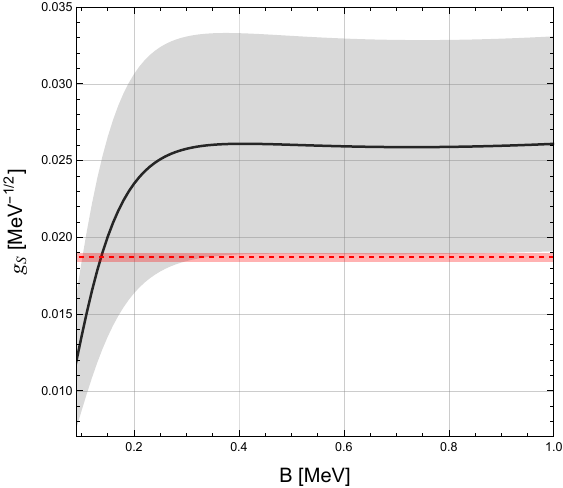}
    \caption{}    
    \label{fig:gSB}
    \end{subfigure}
    \hfill
    \begin{subfigure}{0.48\linewidth}
        \centering
        \includegraphics[width=\linewidth]{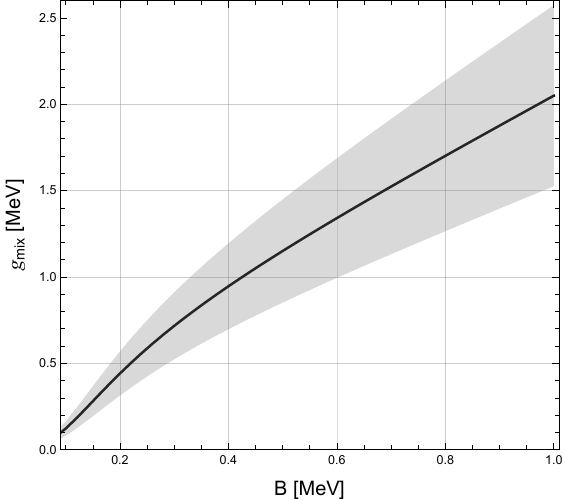}
    \caption{}    
    \label{fig:gmixB}
    \end{subfigure}
    \caption{Behaviour of the couplings $g_S$ and $g_{\text{mix}}$ obtained by fixing the cutoff $\Lambda$ as in Eq.~\eqref{eq:cutoff}. Left: dependence of $g_S$ on the binding energy $B$. The value of $g_S$ extracted from the LHCb fit~\cite{LHCb:2020xds} (see also the discussion in Sec.~\ref{gmixextr}) is shown as a red horizontal line. Right: dependence of $g_{\text{mix}}$ on the binding energy $B$.}
    \label{fig:gsgmixB}
\end{figure}

In the concluding part of this Section, let us reverse the reasoning and start by assuming Eqs.~\eqref{eq:gmixcharged} and~\eqref{eq:cutoff}, the latter fixing the value of $\Lambda$. So $g_{\text{mix}}$ is given by
\begin{equation}
        g_{\text{mix}} = g_S g_T  \frac{m}{\pi^2} \left[
 \text{arctan} \left( \frac{3}{2}\sqrt{\frac{B}{\Delta+|E|}} \right)\sqrt{2m(\Delta+|E|)} -\text{arctan} \left(\frac{3}{2}\sqrt{\frac{B}{|E|}}  \right)\sqrt{2m|E|} \right]
\end{equation}
and substituting back into~\eqref{eq:couplingd}, we get
\begin{equation}
    g^2=8 m m_X^2 g_S^2\left|1-\frac{m}{\pi^2}g_T^2\frac{F(B)}{E+B+i\Gamma_T/2}\right|^2\,,
\end{equation}
where
\begin{equation}
    F(B)=\left[
 \text{arctan} \left( \frac{3}{2}\sqrt{\frac{B}{\Delta+|E|}} \right)\sqrt{2m(\Delta+|E|)} -\text{arctan} \left(\frac{3}{2}\sqrt{\frac{B}{|E|}}  \right)\sqrt{2m|E|} \right]\,.
\end{equation}
Solving the equation $\Gamma_{\rm th} = \Gamma_{\rm exp}$ for $g_S$, rather that for $g_{\text{mix}}$ (see Section~\ref{gmixextr}), we obtain $g_S$ as a function of $B$. The behaviour is shown in Fig.~\ref{fig:gSB}. 

Alternatively, we can also isolate $g_S$ from the loop relation
\begin{equation}
    g_S=\frac{\pi^2}{m g_T}\,\frac{g_{\text{mix}}}{F(B)}\,.
\end{equation}
This allows to bypass the extraction of $g_S$ from LHCb analysis. We substitute this into~\eqref{eq:couplingd}, obtaining
\begin{equation}
    g^2=8 m m_X^2 g_{\text{mix}}^2\left|\frac{\pi^2}{m g_T}\frac{1}{F(B)}-\frac{g_T}{E+B+i\Gamma_T/2}\right|^2\,,
\end{equation}
and then solve $\Gamma_{\rm th} = \Gamma_{\rm exp}$. The dependence of $g_{\text{mix}}$ on $B$ using this alternative method is shown in Fig.~\ref{fig:gmixB}, which is compatible with what has been found in Sec.~\ref{gmixextr} (see Fig.~\ref{fig:gmix2}).

\section{The \texorpdfstring{$Z$}{Z} resonances}
\label{sec:z-res}
The previous model  fits observations remarkably well but it also  raises challenging theoretical questions.
The proposed scheme suggests that the $X(3872)$ is a superposition of a compact isosinglet tetraquark $X_S$ and the neutral component of a ``molecular'' isotriplet which has a uppressed production cross sections. However, the situation could have been
reversed. One might have anticipated the formation of three charged compact tetratquark states, with some mixing of a neutral molecular isosinglet. We propose that the latter case is the one representing the so called $Z$ resonances. 

A set of three particles, $Z^\pm$ and $Z^0$, has been observed in $J/\psi\,\pi$ decays with masses very close to 3872 MeV. These states, generally referred to as $Z(3900)$ (also known as $T_{c\bar{c}1}(3900)$ \cite{ParticleDataGroup:2024cfk}), correspond to $1^{+-}$ resonances, in contrast to the $X(3872)$, which is a $1^{++}$ state. Experimental data indicate that all of these $Z$ states lie slightly above the $D^0\bar{D}^{*0}$ threshold. For this reason, we will   naturally interpret them as $Z_T$ compact tetraquarks.  By analogy, this classification implies the existence of a molecular  state $Z_S$, forming a  (compact, molecular) system $(Z_T, Z_S)$ that mirrors the $(X_S, X_T)$ configuration discussed earlier. In this framework, $Z_S$  corresponds to the molecular counterpart of $Z_T$, although we have reasons to believe that it might not be an allowed state at all.\footnote{The oscillation between a compact tetraquark configuration and a
molecular one can be interpreted as arising from an exchange
interaction, by which a quark $q$ in the diquark is exchanged with the
antiquark $\bar q$ in the antidiquark. Such an exchange interaction clearly preserves charge conjugation $C$, and therefore allows the two configurations to mix. In the diquark-antidiquark picture, in terms of the  spins of the quark pairs, the state corresponding to $X_S$ is 
\be
|X_S\rangle=\tfrac{1}{\sqrt{2}}(|1_{cq},0_{\bar c\bar q}\rangle+|0_{cq},1_{\bar c\bar q}\rangle)
\label{orto}
\ee
with  $C=+1$. This can be Fierz rearranged into %either in $|1_{c\bar c},1_{q\bar q}\rangle_1$ or into
$\tfrac{1}{\sqrt{2}}(|1_{c\bar q},0_{q\bar c}\rangle-|0_{c\bar q},1_{ q\bar c}\rangle)$.
For $q=u$ we have
\be
\tfrac{1}{\sqrt{2}}(\bm D^{*0}\bar D^0-\bar{\bm D}^{*0}D^0)=|X_T\rangle
\ee  
the  $D\bar  D^*$ molecule quantum state with  $C=+1$, i.e. the neutral component of the molecular isotriplet.  
For $q=d$, the state $\tfrac{1}{\sqrt{2}}(\bm D^{*+}\bar D^--\bar{\bm D}^{*-}D^+)$ is obtained. $X_S$ is the  $u\bar u+d\bar d$ isosinglet, and $X_T$ follows, containing both neutral and charged states. 
The  combination orthogonal to~\eqref{orto} is the $C=-1$ state 
\be
\tfrac{1}{\sqrt{2}}(|1_{cq},0_{\bar c\bar q}\rangle-|0_{cq},1_{\bar c\bar q}\rangle)=|Z_T\rangle
\ee
which we associate to the neutral state of the $Z$ triplet of compact tetraquarks. The Fierz rearrangement of $Z_T$ does not contain the $C=-1$ superposition 
$\tfrac{1}{\sqrt{2}}(\bm D^{*0}\bar D^0+\bar{\bm D}^{*0}D^0)$ but only the  $\bm D^{*0}\times \bar{\bm D}^{*0}$ combination (which is $C=-1$ because of the antisimmetry of the cross product).  The latter is too heavy to be the $Z_S$ molecular counterpart. This observation might lead us to suspect that there is no molecular  $Z_S$ and no $(X_S,X_T)\leftrightarrow (Z_T, Z_S)$ mirroring.  Nonetheless we will consider the hypothesis that  a $Z_S$ molecule can mix with $Z_T$ and study its consequences.  
\label{otto}} In addition to the arguments made in footnote~\ref{otto}, assuming a similar near-degeneracy in mass, $Z_S$ would also be located above the threshold, precluding the existence of a loosely bound $Z_S$ state, i.e. of a pole in the low-energy scattering $D\bar D^*$ amplitude at $E=-B$.

If a $Z_S$ molecule were to exist and interact with the $Z_T$ particle, one would anticipate isospin-violating decays of $Z_T$. Specifically, we would expect decays into the $J/\psi\,\eta$ state, which would mirror the isospin-violating pattern observed in the $(X_S,X_T)$ system. However, no such decay has been observed to date. This observation is consistent with the absence of a molecular $Z_S$. Nevertheless, we will conduct further analysis of the case of a $C=-1$, $Z_S$ molecule, and find that isospin violation is not expected in either way.

%This overall pattern suggests a striking structural difference between the two multiplets: the singlet appears as a compact configuration for $1^{++}$ states, whereas it is the triplet to be  compact form for $1^{+-}$ states.\\

Let us then introduce a molecular $Z_S$ state, i.e.  assume that the difference in mass between the compact $Z_T$ state and the molecular $Z_S$ state is such that the mass of the latter falls below the open-charm threshold. 

\begin{figure}[t]
\centering

% --- Diagramma (a) ---
\begin{subfigure}[b]{0.45\textwidth}
\centering
\begin{tikzpicture}
  \begin{feynman}
    \vertex (blob) at (2.2,0) {} ;
    \vertex (v1) at (4.4,0) ;
    \vertex at (3.5,0.3) {\(Z_T\)};
    \vertex (f1) at (6,1) {$J/\psi$};
    \vertex (f2) at (5.8,-1) {\(\pi\)};

    \diagram*{
      (blob) --  (v1),
      (v1) -- [solid] (f1),
      (v1) -- (f2),
    };
  \end{feynman}
\end{tikzpicture}
\caption*{(a)}
\end{subfigure}
%\hfill
% --- Diagramma (b) ---
\begin{subfigure}[b]{0.45\textwidth}
\centering
\begin{tikzpicture}
  \begin{feynman}
    \vertex (i1) at (0,0);
    \vertex at (0.9,0.3) {\(Z_T\)};
%    \vertex [blob, minimum size=1.2cm, fill=gray!20] (blob) at (2.2,0) {};
  \shade[ball color=gray!50] (2.2,0) circle (0.18);
    \vertex at (2.2,0.6) {\(g^Z_{\text{mix}}\)};
    \vertex (v1) at (4.4,0);
    \vertex at (3.5,0.3) {\(Z_S\)};
    \vertex (f1) at (6,1) {$J/\psi$};
    \vertex (f2) at (5.8,-1) {\(\eta\)};    
    \diagram*{
      (i1) -- (blob),
      (blob) -- [dashed] (v1),
      (v1) -- [solid] (f1),
      (v1) -- (f2),
    };
  \end{feynman}
\end{tikzpicture}
\caption*{(b)}
\end{subfigure}

\vspace{0.5em}
\caption{Decay mechanisms for \(Z(3900) \rightarrow J/\psi\, P\) with \(P = \pi,\, \eta\). (a) Isospin-conserving decay. (b) Isospin-violating decay induced by mixing with the isosinglet.}
\label{fig:Z_T_decay}
\end{figure}
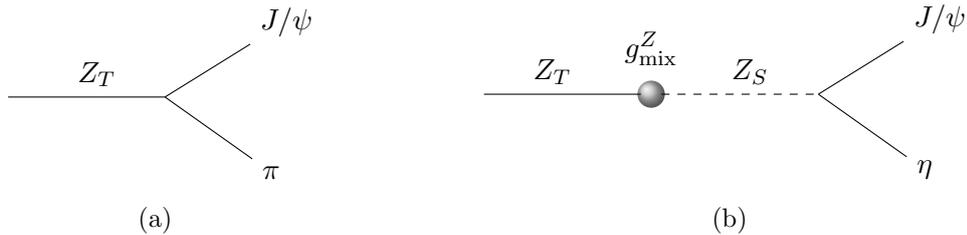

Then, as discussed before, the loosely bound molecule $Z_S$ plays a role in the isospin violating decays of its neutral iso-triplet partner $Z_T$. In this picture, the $Z_S$ particle can appear only in the mixing  represented in Fig~\ref{fig:Z_T_decay}$(b)$.
As we said before, to this day there is no evidence for isospin violating decays for the $Z(3900)$. Nevertheless, an upper bound for the ratio ${\cal I}_Z$ was measured \cite{BESIII:2015aym}
\begin{equation}
   {\cal I}_Z= \frac{{\text{Br}}(Z(3900)\to J/\psi\,\eta)}{{\text{Br}}(Z(3900)\to J/\psi\,\pi)}<0.15\quad \text{[BESIII]}\,.
    \label{eq:br_ratio2}
\end{equation}
We have to ascertain if a $Z_S$ below the $D^{+}D^{*-}$ threshold is in agreement with this upper bound.

Assuming that the mixing between $Z_S$ and $Z_T$ is induced by a $DD^*$ meson loop like in the case of the $X_S-X_T$ oscillation, we have similarly 
\begin{multline}
g_{\text{mix}}^Z= 
g^Z_T\,g^Z_S  \frac{m}{\pi^2} \left[
 \sqrt{-2mE_Z}\,\text{arctan} \left( \frac{\Lambda_Z}{\sqrt{-2mE_Z}} \right)+\right.\\\left.-\sqrt{-2m(E_Z+\Delta)}\,\text{arctan} \left( \frac{\Lambda_Z}{\sqrt{-2m(E_Z+\Delta)}} \right) \right]
\label{eq:gmixz}
\end{multline} 
where $E_Z$ is\footnote{Since the $Z_S$ in the molecular picture is most plausibly composed of the charged mesons, in this case we measure all energies with respect to the charged threshold.}
\begin{equation}
    E_Z=m_Z-m_{D^{+*}}-m_{D^-} = 7.3 \pm 2.6 \,\text{MeV}
\end{equation}
In what follows, for simplicity, we will use the central value.
This means that $\sqrt{-2mE_Z}$ ($\sqrt{-2m(E_Z+\Delta)}$) is the square root of a negative number.  
The correct branch required to ensure unitarity and analyticity of the scattering amplitude follows from restoring the $i\epsilon$ prescription of the propagators in Eq.~\eqref{eq:loopEE}
\begin{equation}
    \lim_{\epsilon\to0^+} \sqrt{-2mE_Z - i\epsilon}
    = -i\sqrt{2mE_Z}\,.
\end{equation}
Moreover, using
\begin{equation}
    \arctan(ix) = i\,\operatorname{arctanh}(x)\,,
\end{equation}
we can rewrite the amplitude for $g^Z_{\mathrm{mix}}$ as
\begin{multline}
g_{\text{mix}}^Z= 
g^Z_T\,g^Z_S  \frac{m}{\pi^2} \left[
 \sqrt{2mE_Z}\,\text{arctanh} \left( \frac{\Lambda_Z}{\sqrt{2mE_Z}} \right)+\right.\\\left.-\sqrt{2m(E_Z+\Delta)}\,\text{arctanh} \left( \frac{\Lambda_Z}{\sqrt{2m(E_Z+\Delta)}} \right) \right]
\label{eq:gmixz2}
\end{multline}
$g^Z_T$ and $g^Z_S$ denote the couplings of $Z_T$ and $Z_S$, respectively, to the $DD^*$ mesons.
These couplings are unknown, but can be estimated with an analysis similar to the one done before. As a matter of fact, $X_S$ and $Z_T$ are spin-1 hidden charm tetraquarks with a very similar mass, therefore we make the  assumption that $g^Z_T \sim g_S$. Considering the mass range where $Z_S$ can be regarded as a loosely bound molecule of $DD^*$ charged mesons, we estimate $g^Z_S$ using the Landau coupling~\cite{landau}
\begin{equation}
    \left(g^Z_S\right)^2=\frac{2\pi}{m}\sqrt{\frac{2B_Z}{m}}
\end{equation}
where $B_Z$ is the binding energy of the $Z_S$ molecule. Then $\Lambda_Z$ can be estimated as in Eq.~\eqref{eq:cutoff}
\begin{equation}
\Lambda_Z = \frac{3}{2}\sqrt{2m B_Z}\,.
\end{equation}
Therefore, we get an expression of $g^Z_\mathrm{mix}$ as a function of $B_Z$
\begin{equation}
g^Z_\mathrm{mix} 
= g_S\,\sqrt{\frac{2\pi}{m}}\left(\frac{2B_Z}{m}\right)^{1/4}\frac{m}{\pi^2}\,F_Z(B_Z)\,,
\label{eq:gmixzewf2}
\end{equation}
where
\begin{equation}
    F_Z(B)=\left[
 \sqrt{2mE_Z}\,\text{arctanh} \left(\frac{3}{2}\sqrt{\frac{B_Z}{E_Z}} \right)-\sqrt{2m(E_Z+\Delta)}\,\text{arctanh} \left( \frac{3}{2}\sqrt{\frac{B_Z}{E_Z+\Delta}} \right) \right]\,.
\end{equation}
In analogy with the $X_T$ case, we consider $B_Z$ in the range $[0.01,1]\,\text{MeV}$ and find the corresponding $g^Z_\mathrm{mix}$.

At present, there are no experimental data available for the individual branching ratios of the decays of the $Z(3900)$. However the following ratio was measured \cite{BESIII:2013qmu}
\begin{equation}
   {\cal R}= \frac{{\text{Br}}(Z^{\pm}\to (D\bar{D}^{*})^{\pm})}{{\text{Br}}(Z^{\pm}\to J/\psi\,\pi^{\pm})}=6.2\pm1.1\pm2.7 \quad \text{[BESIII]}\,.
    \label{eq:br_ratioZ}
\end{equation}
The decay width for $Z_T^{+} \to J/\psi \,\pi^{+}$ is\footnote{The amplitude for the tree-level process $Z_T^{+}\to J/\psi\, \pi^{+}$ is given by 
\begin{equation}
    \mathcal{A}_{ZJ/\psi\pi^+}=G_\pi\,\varepsilon^{\mu\nu\rho\sigma} e_{_{(Z)\mu}}e_{_{(J/\psi) \nu}}^*\,p_{_{(Z),\rho}} p_{_{(\pi),\rho}} \,.
\end{equation}
}
\begin{equation}
\Gamma(Z_T^{+}\to J/\psi\, \pi^{+}) 
= \frac{G_\pi^2}{12\pi} \,p(m^2_Z,m^2_{\pi^+},m^2_{J/\psi})^3\,.
\label{eq:decayZpicharged}
\end{equation}
In analogy with \cite{Esposito:2025hlp} (see Sec.(5)), the width $\Gamma(Z_T^{+} \to D^0 \,D^{*+})$ can be derived fromt the matrix element
\begin{equation}
\langle D^{0} D^{*+} | Z \rangle = g \, e_{_{(Z)}} \cdot e_{_{(D^{*+})}}^{*}\,,
\end{equation}
where
\begin{equation}
g^2 = 8 m m_Z^{2} \, (g^Z_T)^2 \simeq  8 m m_Z^{2} \, g_S^2\,.
\end{equation}
This gives\footnote{The sum over polarizations is
\begin{equation}
\frac{1}{3} \sum_{\text{pol.}} \left| e_{_{(Z)}} \cdot e_{_{(D^{*+})}}^{*} \right|^2 \simeq 1\,.
\end{equation}}
\begin{equation}
    \Gamma(Z_T^{+} \to D^0  D^{*+}) 
= \frac{p(m^2_Z,m^2_D,m^2_{D^{*+}})}{8 \pi m^2_Z} g^2 =\frac{p(m^2_Z,m^2_D,m^2_{D^{*+}})}{\pi} m\,g_S^2 \,.
\end{equation}
Then from Eq. \eqref{eq:br_ratioZ}
\begin{equation}
    G_\pi = \sqrt{\frac{\,p(m^2_Z,m^2_D,m^2_{D^{*+}})\,12\,m\, g_S^2}{p(m^2_Z,m^2_{\pi^+},m^2_{J/\psi})^3 \,\cal{R}}} = (5.4 \pm 1.4) \times 10^{-4} \,\mathrm{MeV^{-1}}\,.
    \label{eq:Gpi}
\end{equation}
We need also the coupling $G_\eta$ between the molecular $Z_S$ and $J/\psi\, \eta$. We have no experimental data for the process $Z_S \to J/\psi \,\eta$, because the $Z_S$ production is very much suppressed to begin with.

In this case we cannot proceed as for the $X_T$ and estimate the coupling through VMD. Therefore we introduce a method to obtain an upper limit on $G_\eta$. As a first step, we assume, as for the molecular $X_T$, that $\Gamma_Z \simeq \Gamma_{D^{*\pm}} = 0.0834 \pm 0.0018 \,\text{MeV}$ \cite{ParticleDataGroup:2024cfk}. Then, we can certainly write
\begin{equation}
    \Gamma(Z_S\to J/\psi\, \eta)=\lambda_Z\,  \Gamma_Z
    \label{eq:Gamma_rho_Z}
\end{equation}
where $\lambda_Z$ is nothing but the branching fraction $\text{Br}(Z_S\to J/\psi\,\eta)$, which gives the probability that the two charm quarks within the molecule are located at a relative distance compatible with the spatial size of the $J/\psi$, hadronize into it, and that simultaneously the two light quarks hadronize into an $\eta$. Therefore, we rewrite $\lambda_Z$ as
\begin{equation}
    \lambda_Z=\mathcal{P}_{c\bar{c}\to J\psi} \times \mathcal{P}_{q\bar{q}\to\eta}
    \label{landa}
\end{equation}
Since we are interested in the compatibility with the upper bound in Eq.~\eqref{eq:br_ratio2}, we estimate an upper limit for $\lambda_Z$ assuming $\mathcal{P}_{q\bar{q}\to\eta}=1$
\begin{equation}
    \lambda_Z < \lambda_{Z,\text{max}}=\mathcal{P}_{c\bar{c}\to J\psi}\,.
\end{equation}
We evaluate the probability that the two charm quarks in the $Z_S$ molecule hadronize into a $J/\psi$ as the overlap integral between the $J/\psi$ and $Z_S$ wavefunctions,
\begin{equation}
 {\cal P}_{c\bar c\to J/\psi} = \int_0^\infty dr\, 4\pi r^2\, \Psi_{_{Z_S}}(r)\, \Psi_{_{J/\psi}}(r)\,,
\end{equation}
where $\Psi_{_{J/\psi}}(r)$ is the $J/\psi$ wavefunction \cite{Grinstein:2024rcu,Germani:2025mos}, and $\Psi_{Z_S}(r)$ is the universal wavefunction for a loosely bound state~\cite{Braaten:2003he,jackiw}
\begin{equation}
\Psi_{_{Z_S}}(r)=\left(\frac{2m B_Z}{4\pi^2}\right)^{1/4}
\frac{\exp(-r\sqrt{2mB_Z})}{r}\,.
\end{equation}
On the other hand, the width of the process $Z_S\to J/\psi\,\eta$ is given by
\begin{equation}
    \Gamma(Z_S\to J/\psi\, \eta) 
    = \frac{G_\eta^2}{12\pi}\, p(m^2_Z,m^2_\eta,m^2_{J/\psi})^3\,.
    \label{eq:Gamma_Z_eta}
\end{equation}
By equating Eq.~\eqref{eq:Gamma_Z_eta} with Eq.~\eqref{eq:Gamma_rho_Z}, and in particular taking the upper bound for $\lambda_Z$, we obtain an upper bound for $G_\eta$:
\begin{equation}
G_{\eta,\text{max}} = 
\sqrt{\frac{12 \pi\,\lambda_{Z,\text{max}}\,  \Gamma_Z}
{p(m^2_Z,m^2_\eta,m^2_{J/\psi})^3}}\,.
\label{G_eta}    
\end{equation}

\begin{figure}[t]
    \centering
    \includegraphics[width=0.6\linewidth]{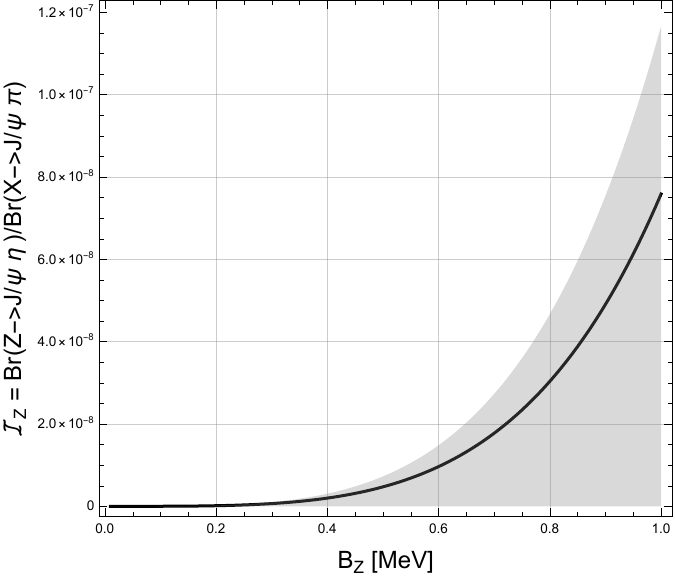}
    \caption{${\cal I}_{Z,\text{max}}$ as a function of  $B_Z$. The black line represents the central value of ${\cal I}_{Z,\text{max}}$ and the gray region contains the values within one sigma.}
    \label{fig:isoz}
\end{figure}

The process $Z_T \to J/\psi \,\eta$ in our model proceeds through the mixing $Z_T-Z_S$. Hence the width for this decay is
\begin{equation}
\Gamma(Z_T\to J/\psi\, \eta) 
= \bigg|\dfrac{g^Z_\text{mix}}{E_Z + B_Z+\frac{i\Gamma_Z}{2}}\bigg|^{2}\frac{G_\eta^2}{12\pi}\, p(m^2_Z,m^2_\eta,m^2_{J/\psi})^3\,.
\label{eq:decayZeta}
\end{equation}
Therefore we arrive at the following expression for the upper bound on the $\mathcal I_Z$ ratio
\begin{equation}
   {\cal I}_{Z,\text{max}} = \frac{{\text{Br}}(Z(3900)\to J/\psi\,\eta)}{{\text{Br}}(Z(3900)\to J/\psi\,\pi)}= \frac{p(m^2_Z,m^2_\eta,m^2_{J/\psi})^3}{p(m^2_Z,m^2_{\pi^0},m^2_{J/\psi})^3}\,\,\bigg|\dfrac{g^Z_\text{mix}}{E_Z + B_Z+\frac{i\Gamma_Z}{2}}\,\, \bigg|^{2}\frac{G_{\eta,\text{max}}^2}{G_\pi^2}\,.
    \label{eq:br_ratio1}
\end{equation}
Using the values for $G_\pi$ and $G_{\eta,\text{max}}$ found  in Eqs.~\eqref{eq:Gpi} and~\eqref{G_eta}, and $g^Z_{\rm mix}$ in Eq. \eqref{eq:gmixz}, we obtain the behavior shown in Fig.~\ref{fig:isoz}.
The estimate we found for the upper bound ${\cal I}_{Z,\text{max}}$ is way below the upper bound reported in Eq.~\eqref{eq:br_ratio2}. This is in agreement with the fact that no isospin violation decay has been ever detected for $Z(3900)$.

\section{Conclusions and outlook}
\label{conclusions}
Based on previous work~\cite{Bignamini:2009sk, Artoisenet:2009wk, Esposito:2013ada, Esposito:2015fsa, Esposito:2020ywk}, we postulate
that at collider experiments, particularly in prompt hadron collisions and $B$ decays, only compact tetraquarks are formed.~\footnote{In  Refs.~\cite{Germani:2025mos,Brambilla:2024thx} the state of the compact tetraquark is taken to be $|(c\bar{c})_{8_c}(q\bar{q})_{8_c}\rangle_{1_c}$. Using a Born-Oppenheimer framework, Ref.~\cite{Brambilla:2026ujo}, argues that the production of the $X(3872)$ proceeds from a color-octet $c\bar{c}$ core.} The $X(3872)$ is interpreted as a mixed state containing a compact isosinglet tetraquark, which we designate as $X_S$, and a neutral component of an isotriplet of molecular states $D\bar D^*$ that we designate as $X_T$. Our postulate is then that the production via hadronization of the $X_T$ and its charged counterparts is highly suppressed in prompt collisions and $B$ decays. However, it still can mix with the compact $X_S$, whose yield is very high. It is this mixing that generates the isospin violations observed in $X(3872)$ decays. The mixing arises from virtual contributions of $DD^*$ pairs to the self energy, and breaks isospin via the neutral mass difference between the neutral and charged pairs. The coupling of $X_T$ to the $DD^*$ pairs can be estimated using the Landau coupling~\cite{landau} of a molecular state to its components.  We have shown that this model is quantitatively viable and self-consistent.  

The $Z(3900)$ ($T_{c\bar c 1}(3900)$) resonance  resonance, is observed  to form a triplet of states, $Z^{\pm,0}$, with opposite charge conjugation to that of the $X(3872)$. In our interpretation, these states, found slightly above threshold, are all compact tetraquarks. A neutral isosinglet partner, $Z_S$, of molecular nature, might also contribute to the $Z_T$ decay in the mixing amplitude, although we observe that it might not be allowed at all. We would have then two mirror multiplets: $(X_S, X_T)$, (compact, molecular), and $(Z_T, Z_S)$, (compact, molecular). In either case we predict no substantial isospin violations in $Z^0(3900)$ decays.

Thus, the model presented solves at the same time the problem of isospin violations in $X$ and $Z$ decays, and gives a  picture of the spectroscopy of states, namely, the absence of charged partners of the $X$ and absence of the neutral particles doubling, for both $X$ and $Z$.   

Let us consider, for instance, the isolated state $T_{c\bar c1}(4430)^+$ (also known as $Z(4430)$). This is most likely the radial excitation of $X_S$~\cite{Maiani:2008zz}. In this scheme this is also a compact tetraquark, and its quark rearrangements cannot match any meson thresholds to allow molecular states. 

We also recall that, to date, there is no evidence of a $X$-like particle in the beauty sector. We might speculate that in the $b$-sector all molecular quark arrangements are not visible, as it occurs for charmed $X_T$, and the compact $X_S^{(b)}$ happens to be sufficiently heavier and strongly coupled to the threshold to get significantly broadened by $B\bar B^*$ decay processes, making it particularly broad.  

In~\cite{Zhang:2024fxy}, isovector molecular partners of the $X(3872)$ (a molecule in that scheme) are also discussed. In that study, these states are identified as virtual state poles of the scattering matrix, which explains, in a different way, why they are not observed.

Various extensions of the minimal model presented herein can be conceived too. For instance, a partial compositeness of the states could be introduced, as discussed in~\cite{Esposito:2025hlp}. This topic will be the subject of further study, in conjunction with a re-examination of the known spectroscopy along the lines outlined herein.  

\acknowledgments
The work of B.G. is
supported in part by the U.S. Department of Energy under grant number DE-SC0009919. 

\newpage
\bibliographystyle{JHEP}
{\footnotesize
\bibliography{biblio}}

\end{document}